\documentclass[preprint,12pt]{elsarticle}

\usepackage{natbib}
\usepackage{xcolor}  
\usepackage{ulem}

\usepackage{graphicx}    
\usepackage[utf8]{inputenc}
\usepackage{amssymb}
\usepackage{amsmath}
\usepackage{hyperref} 
\usepackage{cleveref}
\usepackage{orcidlink}

\journal{Nuclear Physics B}

\begin{document}

\begin{frontmatter}



\title{Locomotion in CAVE: Enhancing Immersion through Full-Body Motion}


\author[1]{Xiaohui Li\orcidlink{0009-0003-5988-1340}}
\author[2]{Xiaolong Liu\corref{cor1}}
\author[2]{Zhongchen Shi\orcidlink{0000-0003-1296-8564}\corref{cor1}}
\author[2]{Wei Chen\orcidlink{0000-0001-6314-5600}}
\author[2]{Liang Xie\orcidlink{0000-0002-8286-6785}}
\author[3]{Meng Gai\orcidlink{0009-0004-2088-7937}}
\author[4]{Jun Cao}
\author[1]{Suxia Zhang\orcidlink{0000-0002-1790-7958}}
\author[2]{Erwei Yin\orcidlink{0000-0002-2147-9888}}

\affiliation[1]{
  organization={Tianjin University}, 
  addressline={}, 
  city={Tianjin},
  postcode={300072},
  state={}, 
  country={China}
}

\affiliation[2]{
  organization={Defense Innovation Institute, Academy of Military Sciences},
  addressline={},
  city={Beijing},
  postcode={100091},
  state={},
  country={China}
}

\affiliation[3]{
  organization={Beijing University},
  addressline={},
  city={Beijing},
  postcode={100871},
  state={},
  country={China}
}

\affiliation[4]{
  organization={Nanjing Ruiyue Technology Co., Ltd.},
  addressline={},
  city={Nanjing},
  postcode={210012},
  state={},
  country={China}
}
\begin{abstract}
Cave Automatic Virtual Environment (CAVE) is one of the virtual reality (VR) immersive devices currently used to present virtual environments.  However, the locomotion methods in the CAVE are limited by unnatural interaction methods, severely hindering the user experience and immersion in the CAVE. We proposed a locomotion framework for CAVE environments aimed at enhancing the immersive locomotion experience through optimized human motion recognition technology.  Firstly, we construct a four-sided display CAVE system, then through the dynamic method based on Perspective-n-Point to calibrate the camera, using the obtained camera intrinsics and extrinsic parameters, and an action recognition architecture to get the action category. At last, transform the action category to a graphical workstation that renders display effects on the screen.
We designed a user study to validate the effectiveness of our method. Compared to the traditional methods, our method has significant improvements in realness and self-presence in the virtual environment, effectively reducing motion sickness.
\end{abstract}

\begin{graphicalabstract}
\begin{figure}
    \centering
    \includegraphics[width=1.0\linewidth]{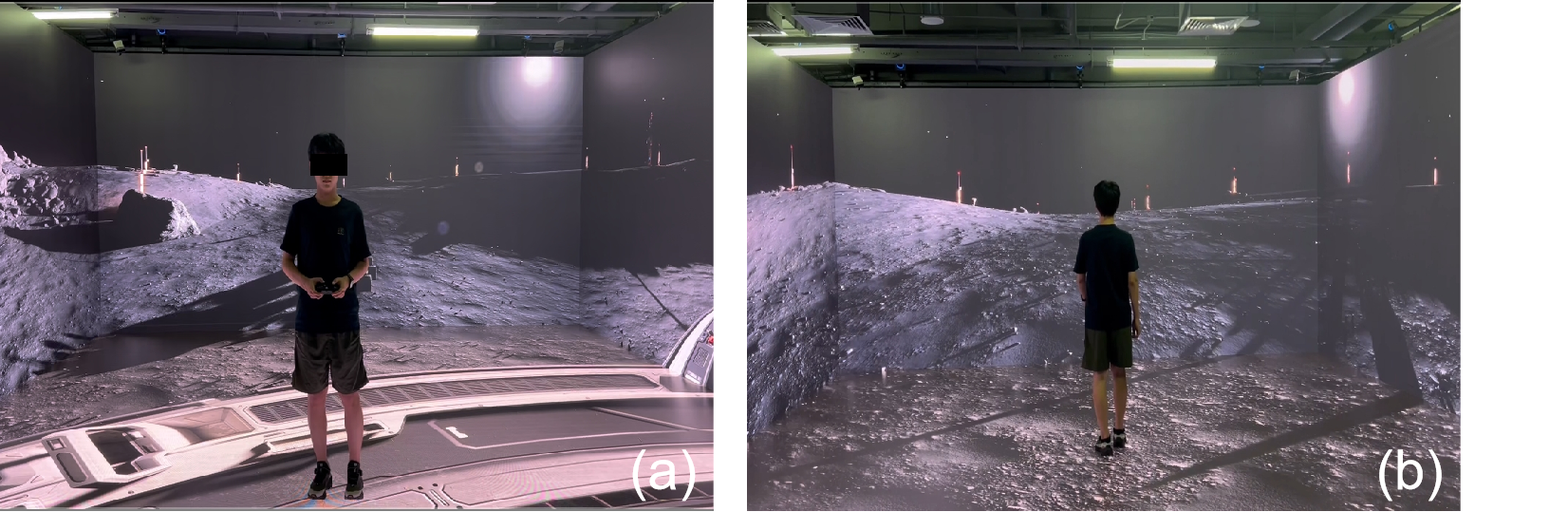}
    \caption{User study scenario. (a) The participant is navigating the scene in  CAVE with controller-based locomotion; (b) the participant is navigating the scene in  CAVE with Embodied CAVE Locomotion.}
    \label{fig:exp}
\end{figure}
\end{graphicalabstract}

\begin{highlights}
\item A pipeline for human motion recognition based on locomotion in CAVE is proposed.
\item We propose a real-time, high-precision human motion recognition algorithm adapted for CAVE locomotion.
\item We design a user study to validate the effectiveness of our proposed method in significantly enhancing the immersive locomotion experience within CAVE environments.
\end{highlights}


\end{frontmatter}



\section{Introduction}
\label{sec1}

Locomotion is a fundamental interaction in virtual reality ($VR$) applications. Currently, most works on $VR$ immersive experiences rely on head-mounted displays (HMDs)\cite {lo2023}\cite{williams2007exploring}. However, wearing an HMD can restrict the user's field of view and cause discomfort after prolonged use, reducing motion fidelity and overall immersion\cite{sharples2008virtual}.

In contrast, CAVE automated virtual environment (CAVE) systems offer a highly immersive motion experience without the need for a head-mounted device. Their presentation methods primarily employ multi-screen or multi-view projector-based presentations\cite{wang2024examining}. This inherent design advantage makes CAVE ideal for applications requiring high levels of immersion, collaboration, and minimal physical restriction. 
However, locomotion in CAVE is limited by unnatural methods, severely impacting user experience and immersion. While traditional input devices, such as mice, keyboards, data gloves or game controllers, can provide a basic locomotion method for CAVE, their unnatural interaction methods can compromise user immersion and potentially cause discomfort or motion sickness\cite{tuena2023}. Furthermore, motion methods based on full-body pose capture have been used in CAVE\cite{ebmer2024real}.
Thus, existing locomotion methods in CAVE are insufficient, limiting their ability to provide immersive experiences.

In this paper, we propose a locomotion framework for CAVE environments aimed at enhancing the immersive locomotion experience through optimized markerless human motion recognition technology. First of all, we construct a four-sided CAVE display system. Next, we dynamically calibrate its cameras using a Perspective-n-Point-based method. The resulting camera intrinsics and extrinsics are then fed into an action recognition architecture to classify actions. Finally, the recognized action category is sent to the graphical workstation to render corresponding display effects.
The contributions of this paper can be summarized as follows:
\begin{itemize}
    \item A pipeline for human motion recognition based on locomotion in CAVE is proposed;
    \item We propose a real-time, high-precision human motion recognition algorithm adapted for CAVE locomotion;
    \item We design a user study to validate the effectiveness of our proposed method in significantly enhancing the immersive locomotion experience within CAVE environments.
\end{itemize}

\section{Related Work}
\label{sec1}
\subsection{CAVE System}
CAVE generally refers to an immersive room-sized environment enclosed by multiple walls, which can be implemented using projection or direct-view technologies such as LED video walls.
For a more comprehensive review of the CAVE methods, we recommend readers read the surveys\cite{muhanna2015virtualSurvey,manjrekar2014caveReview}. 

Over the past three decades, CAVE systems have been categorized into three types based on their display configuration: projection-based CAVE \cite{cruz2023surround}, direct-view CAVE \cite{febretti2013cave2}, and stereoscopic displays\cite{redlarski2018stereoscopy}.
Cruz-Neira et al. \cite{cruz1992cave} first demonstrated a CAVE in 1992, using multiple projectors to project images onto the walls, floor, and ceiling. 
With the continuous advancement of display technology, many CAVE systems rely directly on high-resolution screens to create immersive spaces. Andrew et al. \cite{febretti2013cave2} first proposed a high-resolution LED method. Their work not only demonstrated how high-resolution LED video walls could be seamlessly integrated with the original CAVE concept but also provided a more sophisticated platform for large-scale data visualization and multi-user collaboration. After the advent of stereoscopic display technology, Lukasz et al. \cite{redlarski2018stereoscopy} first incorporated it into CAVEs and studied how to synchronize and calibrate images in multi-screen, multi-computer CAVE systems to ensure accurate stereoscopic rendering. Charilaos et al.\cite{2015}comprised 416 high-density LED-backlit LCD displays, which visualize gigapixel-resolution data while providing 20/20 visual acuity for most of the visualization space. Hikaru et al.\cite{takatori2019large} designed a super large-scale immersive display and make sure that the method can generate the correct perspective from any position inside the screen viewing area. Elisabeth et al.\cite{mayer2024led} designed a system that comprised of 5 borderless LED screens and compared it with a traditional projector-based CAVE by user feedback.
However, these works did not focus on user interaction with the CAVE virtual environment.

\subsection{User interaction within CAVE}
Current works on user interaction within CAVE follow a cyclical path: from the human body itself, to external assistive devices, and then back to the human body. Cruz-Neira et al. \cite{cruz1992cave} placed trackers on the user's head or glasses to continuously capture spatial position and orientation in real time and dynamically adjust the projected image, ensuring that the virtual world remains perspective-correct and visually stable regardless of user movement. Bowman et al. \cite{bowman1999formalizing} developed a tracking wand with integrated trackers and buttons to enable users to actively interact with the CAVE's virtual environment. These devices expanded CAVE's interactive capabilities and quickly became the de facto standard for general, high-performance interaction. He et al. \cite{he2007research} achieved more complex CAVE interactions by using data gloves to track user gestures in real time. However, all of these interaction methods utilize hardware and do not focus on locomotion within the CAVE.

\subsection{Natural walking tracking}
The work by Shin et al.\cite{shin2024implementation} demonstrated the effective implementation and validation of IPL using low-cost depth cameras (Kinect V2) to track user joint positions. Their study successfully established that such vision-based IPL methods can achieve higher user comfort and reduced motion sickness compared to passive controllers. However, their method are designed for HMD contexts and do not address the complex multi-view tracking and dynamic calibration challenges inherent to large-scale CAVE systems.
With advances in computer vision and human motion recognition, methods for achieving natural locomotion within CAVE are gaining increasing attention.  Lu et.al \cite{Lu2025ICMRCAVELomotion}  introduced MoRLACS, a monocular RGBD-based locomotion framework for CAVE systems that improves navigation efficiency and user immersion by combining natural walking with controller-based interaction; however, its limitations include reliance on head tracking for local movement, which may not capture full-body motion and could lead to less precise or immersive navigation compared to full-body tracking systems. 

\section{Method}
\label{sec1}

In this section, we design and implement a CAVE human action recognition system based on a three-layer collaborative architecture, aiming to achieve efficient, accurate, and environmentally adaptable real-time action recognition and posture detection. We describe the proposed technical solution, including the CAVE system(\autoref{sec: CAVE System}), dynamic calibration (\autoref{sec: CAVE Calibration}), action recognition and posture detection algorithms(\autoref{sec:HAR}), and performance optimization strategies(\autoref{sec:transform}).

\subsection{Our CAVE System}
\label{sec: CAVE System}



\begin{figure}[h]
    \centering
    \includegraphics[width=1.0\linewidth]{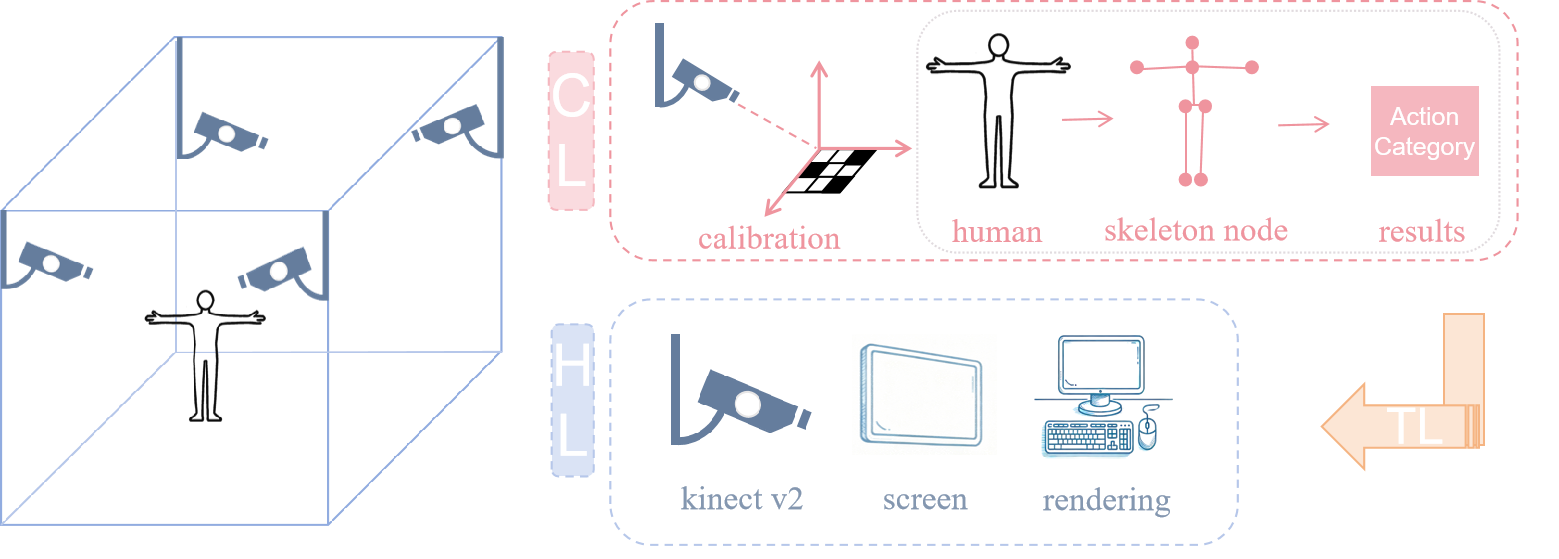}
    \caption{Overall System Architecture: CL denotes the Computing Layer; HL denotes the Hardware Layer; TL denotes the Transmission Layer.} 
    \label{fig:fullsystem}
\end{figure}

In this section, the CAVE system features a three-tier collaborative architecture, consisting of a hardware layer (\autoref{fig:fullsystem} $HL$), a computation layer with GPU servers for data processing(\autoref{fig:fullsystem} $CL$), and an information transmission layer (\autoref{fig:fullsystem} $TL$) utilizing high-speed optical fiber and $UDP$ for efficient, high-precision performance.

At the $HL$, we built a platform consisting of a multi-dimensional environment simulation system and an action recognition system. The former utilizes four ultra-high-resolution $LED$ array screens to create an immersive display space. Driven by $4$ high-performance graphics workstations in the computing layer, it generates realistic virtual scenes and dynamic calibration patterns. The latter comprises a multimodal perception matrix consisting of four depth cameras. This system is capable of synchronously capturing video streams, tracking dynamic objects, and obtaining precise $3D$ motion data of key points on the human body with submillimeter accuracy.

The $CL$, the core of the system, features four GPU servers, a camera calibration module, and a motion recognition system. These systems are responsible for real-time pre-processing, intelligent decision-making, and analysis of the raw data transmitted from the perception layer.

The $TL$ uses high-speed fiber optic networks and $UDP$ (User Datagram Protocol) protocols to support efficient data interaction of the action recognition results from the computing layer to the hardware layer, supporting the real-time performance of the system.


\subsection{Dynamic Self-Adaptive Calibration}
\label{sec: CAVE Calibration}

To overcome the limitations of traditional physical calibration methods, we design a dynamic self-adaptive calibration scheme based on the  Perspective-n-Point(PnP) algorithm. Our method can achieve efficient, precise, and environmentally adaptive acquisition of camera parameters without physical calibration boards. 

\begin{figure}[h]
    \centering
    \includegraphics[width=0.5\linewidth]{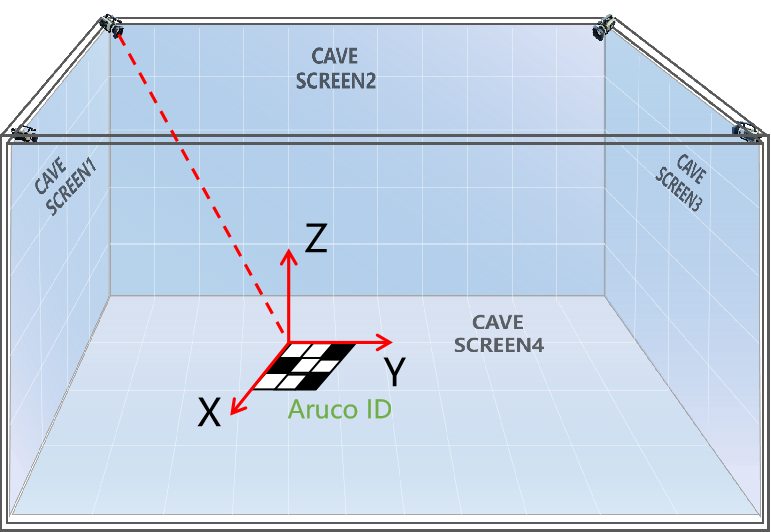}
    \caption{The X-Y-Z coordinate system is generated from the corner points of the Aruco markers.} \label{fig:figbiaoding}
\end{figure}

When the calibration process begins to initiate, the four CAVE screens synchronously display pre-designed Aruco code patterns with known 3D world coordinates as shown in \autoref{fig:figbiaoding}. Subsequently, the system detects and extracts the 2D projection points of these Aruco codes in the camera images through high-precision image processing techniques. By combining the 2D projection points with their corresponding 3D spatial coordinates, the $PnP$ algorithm is used to solve for each camera, obtaining its extrinsic parameters and intrinsic parameters. To ensure millimeter-level precision, we use parameter initialization based on prior knowledge and call the OpenCV library for precise calculation. Finally, check the results using re-projection validation and by visualizing the spatial coordinate axes. In general, our method reduces calibration time from hours to minutes, establishing a reliable spatial reference for high-precision action recognition.

\subsection{Real-Time Action Recognition and Orientation Detection System}
\label{sec:HAR}

\begin{figure}[h]
    \centering
    \includegraphics[width=1.0\linewidth]{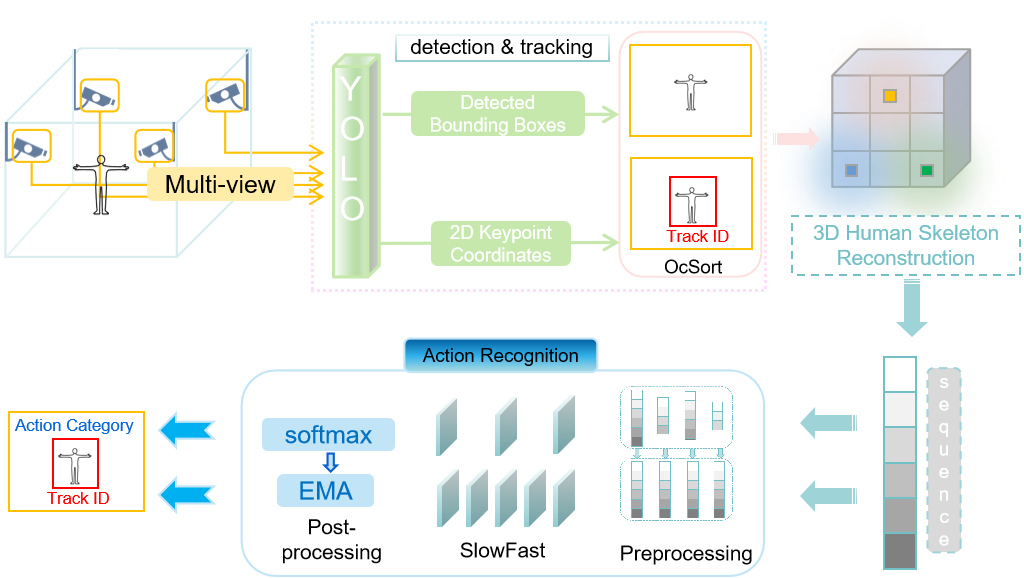}
    \caption{Overview of the Real-Time Action Recognition and Orientation Detection System framework. The system first performs human detection and tracking on the images, then reconstructs 3D skeleton sequences from the 2D keypoint coordinates. Finally, a high-performance action recognition network analyzes these 3D skeleton sequences and outputs the recognized action category and user ID.}
    \label{fig:har}
\end{figure}

Our method aims to develop an efficient, real-time CAVE-based human action recognition framework that enables natural interaction with a virtual environment. \autoref{fig:har} shows that our framework is a multi-stage pipeline including three core modules: 2D human detection and tracking, 3D human skeleton reconstruction, and action recognition.

Our system is established within the above CAVE environment, where multiple cameras installed from different perspectives capture video sequences of the user. These video streams serve as the input for our action recognition framework. The framework first performs human detection and tracking on the multi-view images before reconstructing 3D skeleton sequences from the 2D keypoint coordinates. Finally, a high-performance action recognition network analyzes the 3D skeleton sequences, outputting the recognized action category and user ID, which then drives the CAVE's graphical workstation to render corresponding virtual effects in real time.

To accurately detect and track humans from multiple viewpoints, we employ a two-stage 2D processing pipeline. First, we utilize the lightweight and efficient YOLO model to process video frames from one of the four cameras in real time, detecting bounding boxes and extracting 2D keypoint coordinates. Subsequently, we feed the YOLO output (bounding boxes and keypoints) into the OcSort tracking algorithm. OcSort is an occlusion-aware tracker that can robustly handle occlusion issues and assign a unique track ID to each detected person. This step ensures that we can accurately associate and track the movement trajectories of specific users in multi-view and multi-user scenarios.

To obtain more descriptive motion features from the 2D images, we perform 3D skeleton reconstruction for each tracked person. This module takes the 2D keypoint coordinates with their associated track IDs from OcSort as input. Using multi-view geometry principles and triangulation methods, it projects the 2D keypoints of the same user from different camera viewpoints into 3D space, thereby reconstructing a complete 3D human skeleton. The reconstructed skeleton data is stored as a continuous sequence, providing high-dimensional motion information for subsequent action recognition.

Action recognition is the core of this framework, aiming to classify the 3D skeleton sequences into predefined action categories. This module includes three sub-steps: preprocessing, the SlowFast network architecture, and post-processing. Before inputting the 3D skeleton sequences into the model, we preprocess them to eliminate noise and normalize the data. This may include translation and rotation normalization of the skeletons to ensure the model is not affected by the user's initial position and orientation. We adopt the SlowFast network as the core model for action recognition. SlowFast is a dual-path model in which a "slow path" processes input sequences at a low frame rate to capture spatial information, while a "fast path" captures rapidly changing motion information at a high frame rate. This design enables it to capture both the static appearance and dynamic changes of an action, thereby achieving high-precision action recognition. The output of the SlowFast model is passed through a softmax layer, yielding a probability distribution for each action category. To improve the stability and robustness of the classification, we further use an Exponential Moving Average (EMA) algorithm to smooth the action probabilities over consecutive frames. Finally, the model outputs a definitive action category and the corresponding track ID, which is then sent to the CAVE's graphical workstation for real-time rendering.

\begin{figure}[h]
    \centering
    \includegraphics[width=1.0\linewidth]{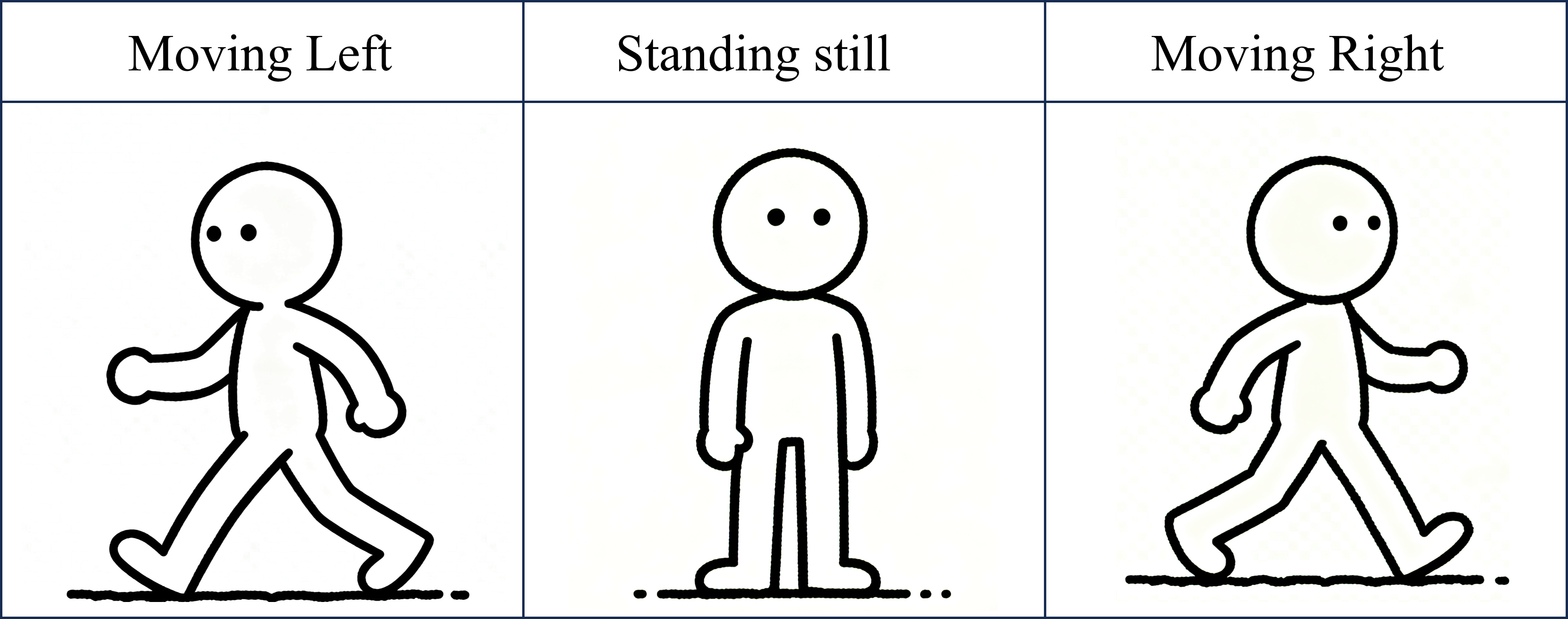}
    \caption{The actions in the dataset: moving left, standing still, and moving right.}
    \label{fig:action}
\end{figure}

To improve the recognition accuracy, we established a dedicated interaction action dataset for the CAVE system. During the data collection, the participants stood in the center and performed the actions, while four cameras captured the data synchronously. The dataset contains a total of 12,000 samples, each lasting 1-2 seconds and including a complete action. 
\autoref{fig:action} shows three actions in the dataset. To facilitate locomotion in the CAVE, the dataset includes four corresponding actions: in-place stepping (left, right, and forward), which map to the virtual movements of moving left, moving right, and moving forward; and random stationary movements for the stand still command.
The uniformly preprocessed dataset was used to train the SlowFast model. The trained model achieved an accuracy of 79.0\% on the Kinetics dataset.

\subsection{Performance Optimization and Technical Innovations}
\label{sec:transform}

The transport layer adopts a multi-threaded asynchronous pipeline processing, allowing the main stages of video acquisition, posture detection, tracking, and action recognition to be executed asynchronously and in parallel, thereby maximizing hardware utilization. The reconstructed 3D iskeleton data can be provided to both the action recognition and orientation detection modules simultaneously and in parallel, avoiding redundant calculations. Finally, the system can dynamically adjust the processing frame rate based on the current computational load to optimize real-time performance while maintaining recognition accuracy. This method significantly improves the system's robustness and performance in scenarios with high occlusion and high dynamics.

\section{User Study}
This section details the design of a user study conducted within the CAVE environment to evaluate the effects of our proposed method on simulator sickness and perceived task load.

\subsection{User Study Design}
\textbf{Participants.}
The study included 20 participants (12 males, 8 females). Their mean age was $25.7$ years ($SD = 1.95$).Most of the participants had previous virtual reality experiences. 
The participants had normal corrected vision, and none reported visual or physical impairment. 
There is one control condition ($CC$) and one experimental condition ($EC$). $CC$ is the Xbox controller method. $EC$ is our method.

On the first day of the experiment, all participants navigated the CAVE using a handheld controller and completed a series of questionnaires both before and after the experiment. To ensure that all participants were in a well-rested state and to mitigate any potential crossover effects between the two interaction methods, the experiment was conducted on different days. To mitigate potential ordering effects, we employed a counterbalanced design. The 20 participants were split equally: 10 completed the CC first, followed by the EC two days later, and the remaining 10 completed the EC first, followed by the CC two days later.

\textbf{Hardware and software setup.}
The hardware platform for the experiment comprised a group of graphics workstations, a Microsoft Kinect V2 camera, and a standard Xbox controller. The workstation was equipped with an NVIDIA A5000 GPU. The experimental tasks were programmed using Unreal Engine, achieving the function of Scene Roaming. Subjective evaluations were collected through questionnaires, and data analysis was accomplished using IBM SPSS. The virtual environment was rendered at a rate of 30 frames per second.

\textbf{Task.}
The experiment was conducted in a CAVE environment, situated within a closed, windowless room to eliminate any external light interference. The experimental sessions were uniformly scheduled between 4:00 PM and 5:00 PM to mitigate any potential influence from participants' state of mind, such as grogginess from waking up. The experimental scenario consisted of a real-time rendered virtual environment, and participants were instructed to navigate along a predefined, consistent route. To maximize the sense of immersion, all participants were required to stand in the exact center of the space formed by the four CAVE screens at the start of each session. Before and after each experimental trial, participants completed a subjective questionnaire, including the Simulator Sickness Questionnaire (SSQ)\cite{9090573}, to quantify their current level of simulator sickness. To ensure participants were adequately rested and to prevent any potential cumulative effects of motion sickness, a minimum interval of one day was maintained between each experimental trial.
\begin{figure}
    \centering
    \includegraphics[width=1.0\linewidth]{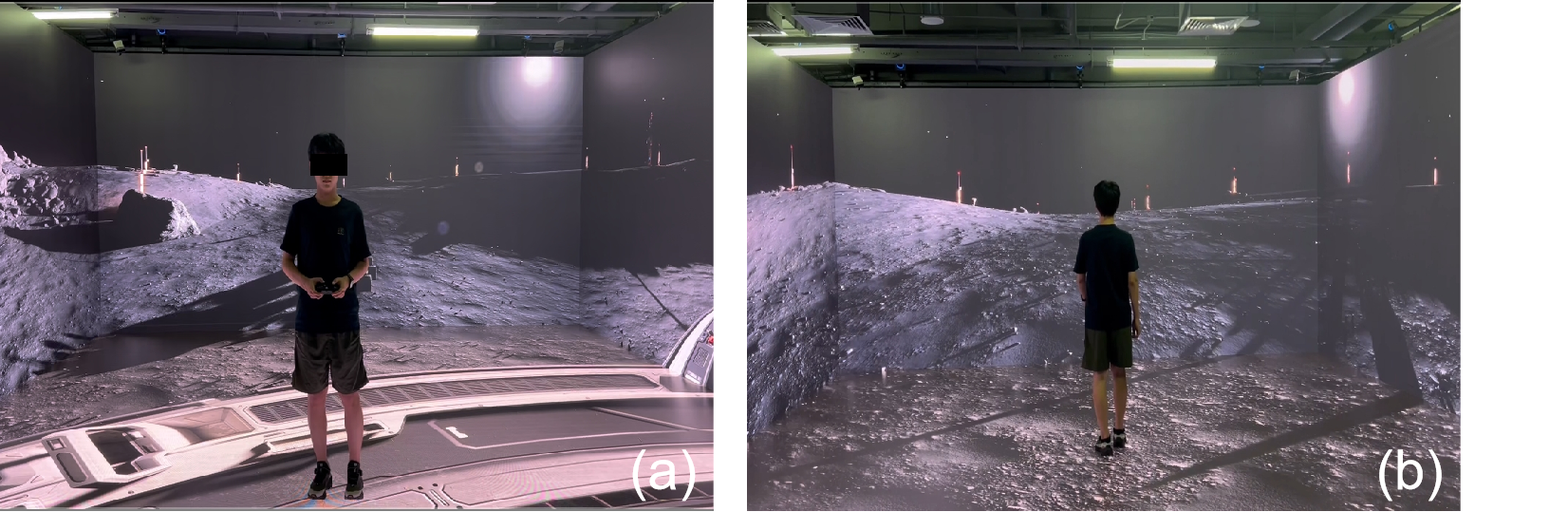}
    \caption{User study scenario. (a) The participant is navigating the scene in  CAVE with controller-based locomotion; (b) the participant is navigating the scene in  CAVE with Embodied CAVE Locomotion.}
    \label{fig:exp}
\end{figure}

\textbf{Procedure.}
Before the experiment, all participants completed their personal information form. Then, a facilitator demonstrated and explained the tasks and objectives for each experimental group, guiding participants on fundamental operations such as direction control, forward, and backward movement using either the controller or gait-based locomotion. Once the formal experiment began, participants were required to maintain focus and complete a pre-defined route from point A to point B within one minute. \autoref{fig:exp} illustrates the user study scenario. The virtual environment was meticulously designed to minimize distractions during locomotion, thereby ensuring the stability and purity of the experimental conditions.

\textbf{Hypotheses.}
In the experiment, each participant participated in two rounds of experiments, respectively: $CC$ and $EC$.

We proposed the following hypotheses:

\textbf{H1}: Compared to  $CC$ , $EC$ alleviates motion sickness.

\textbf{H2}: The task load of $CC$ is higher than that in $EC$. 

\textbf{H3}: The sense of immersion of $EC$ is better than that in $CC$. 

\textbf{Metric.}   
We mainly used the following four subjective questionnaires to count the subjective metrics of the participants: The standard MSSQ questionnaire was completed by the participants before the experiment to record their sensitivity to motion sickness; The standard SSQ questionnaire was divided into SSQ-pre and SSQ-post, which were filled out before and after the experiment respectively to completely record the participants' motion sickness status at that time; The standard IPQ\cite{schubert2001experience} questionnaire and the standard NASA TLX questionnaire \cite{hart1988development,hart2006nasa} was filled out by users after completing each method.

\textbf{Statistical analysis.} 
First of all, we verified the normality assumption of our data distribution using the Shapiro-Wilk\cite{shapiro1965analysis} test, followed by Mauchly's test \cite{mauchly1940significance} to assess the assumption of sphericity. If sphericity was violated, a Greenhouse-Geisser correction was applied to the data. A subsequent overall analysis of variance (ANOVA) was conducted to test the null hypothesis that there were no statistically significant differences between the two cases. For time-dependent variables, we also employed Cohen's d value \cite{cohen2013statistical} to quantify the effect size and qualitatively classified the effect scale into six levels based on the d value range: very large (d>2.0), large (2.0>d>1.2), medium (1.2>d>0.8), small (0.8>d>0.5), very small (0.5>d>0.2), and negligible (0.2>d>0.01). All analyses were performed using SPSS software.

\subsection{Results}

We collected and analyzed data from questionnaires completed by participants after they participated in the experiment. Our experiment compares the impact of controller-based and action-recognition-based locomotion on user simulator sickness in a CAVE environment. 
The overall ANOVA reveals that there is a statistically significant difference between the two conditions for the scene (F(1,19)=18.711, p<0.001, $\eta$ = .496). Due to the non-normal distribution of the paired differences, we employed the Wilcoxon Signed-Rank Test to compare the EC and the CC. The results showed a statistically significant difference in SP scores between the EC  and the CC (Z = -3.722, p < 0.001).

This finding indicates that our proposed locomotion method significantly enhanced the user's sense of self-presence in the virtual environment.

\begin{figure}[h]
    \centering
    \includegraphics[width=1.0\linewidth]{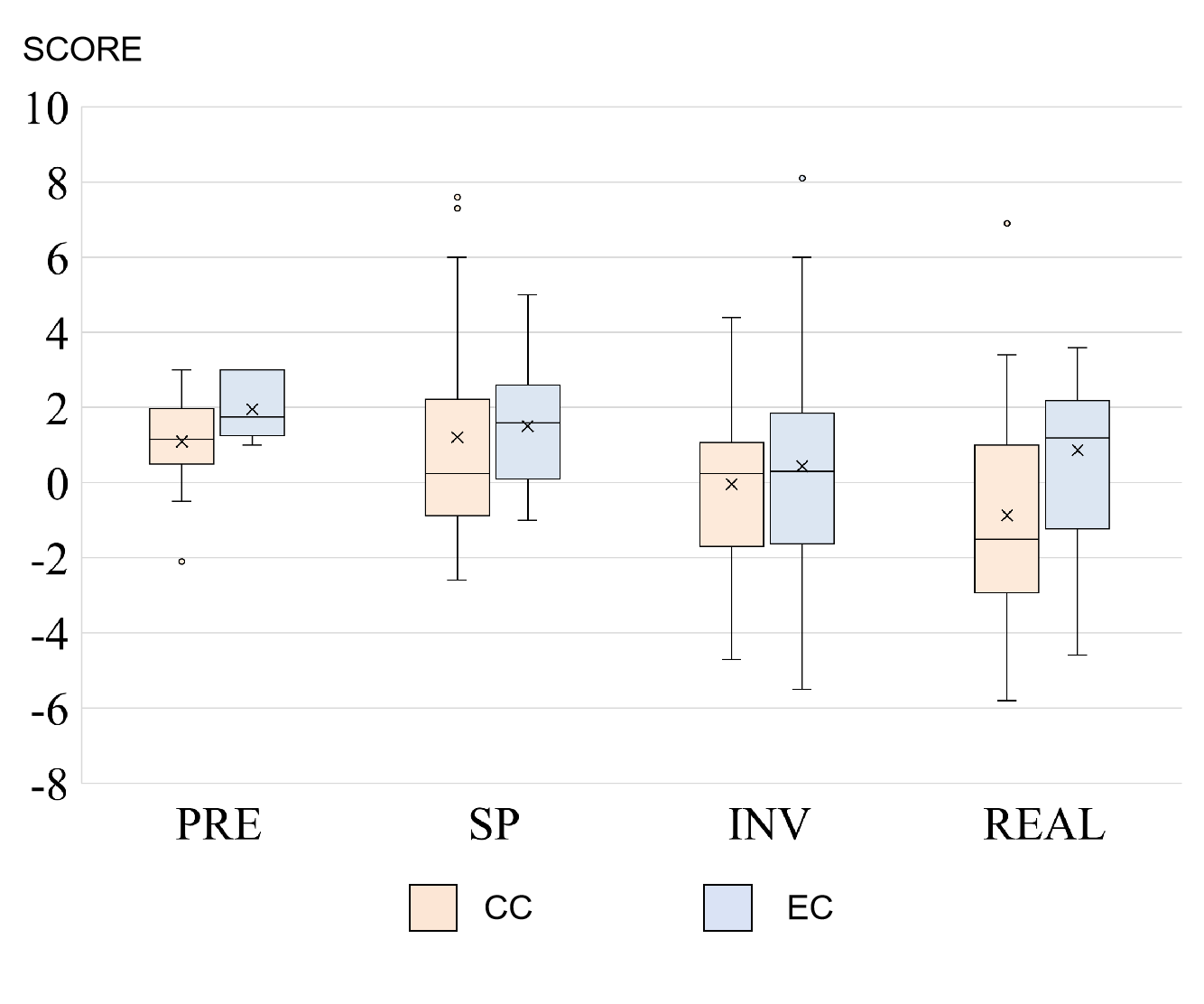}
    \caption{Mean IPQ for PRE, SP, INV, and REAL across each method. The colors respectively represent the controller-based locomotion and Embodied CAVE Locomotion.}
    \label{fig:ipq}
\end{figure}

\textbf{Presence.}
\autoref{fig:ipq} illustrates the distribution of user responses for the group presence questionnaire. 
The results indicate that the participants who used our method reported higher scores for every aspect of presence, self-presence, involvement, and realness. \autoref{fig:ipq} reveals that subjective scores for our method condition consistently trended higher across all questions.

\begin{figure}
    \centering
    \includegraphics[width=1.0\linewidth]{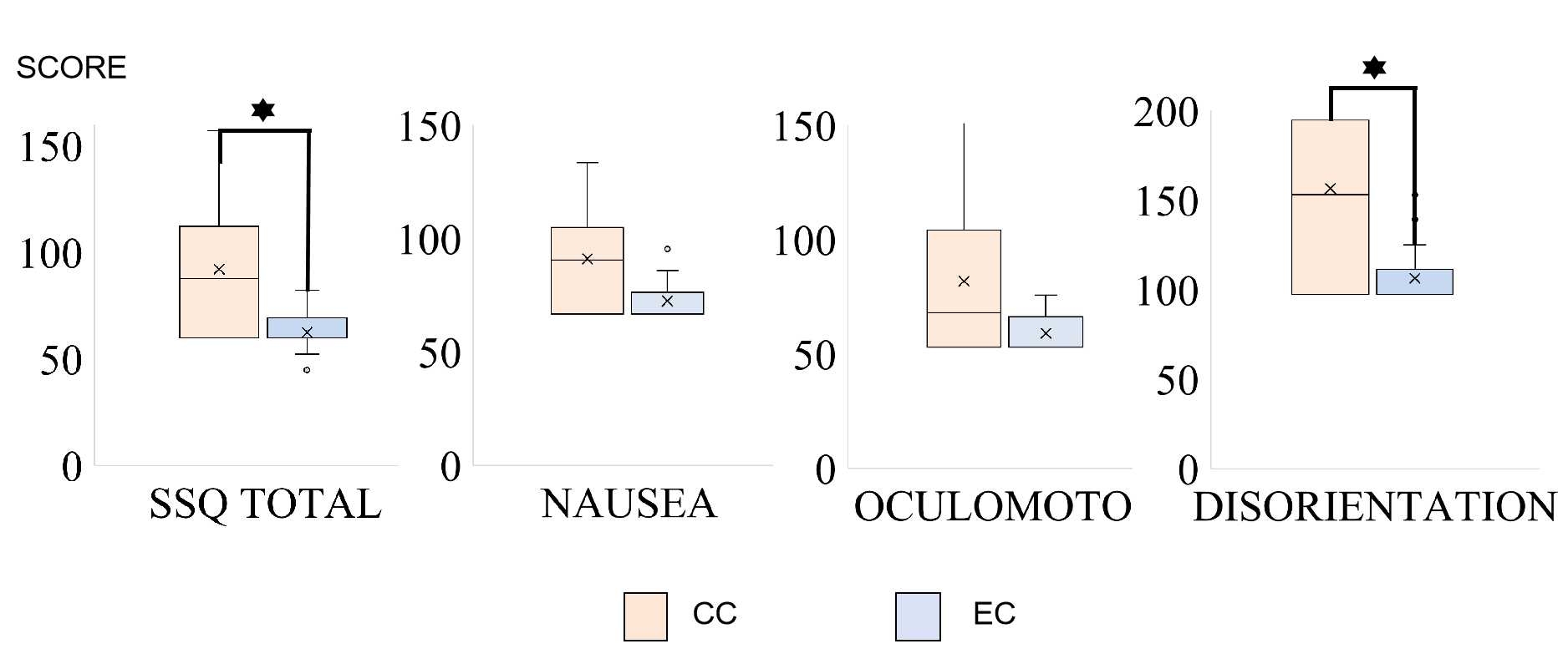}
    \caption{Mean SSQ for Total scores, Nausea, Oculomotor, and Disorientation cross each method. The colors respectively represent the controller-based locomotion and Embodied CAVE Locomotion.}
    \label{fig:ssq}
\end{figure}

\textbf{User Experience.}
In \autoref{fig:ssq}, the SSQ total scores shows a distinct difference between the two control methods. The action-recognition-based control group showed lower scores than the controller-based control group across all metrics, including Total scores, Nausea, Oculomotor, and Disorientation discomfort.
This suggests that the action-recognition method is more effective at alleviating symptoms of simulator sickness.
These results provide further evidence for the superiority of our method in reducing simulator sickness. This indicates that the user experience in terms of simulator sickness was more consistent and stable with the action-recognition control. These findings provide strong evidence that embodied locomotion through action recognition offers a more comfortable and less disorienting immersive experience in CAVE environments.

\begin{figure}
        \centering
        \includegraphics[width=1.0\linewidth]{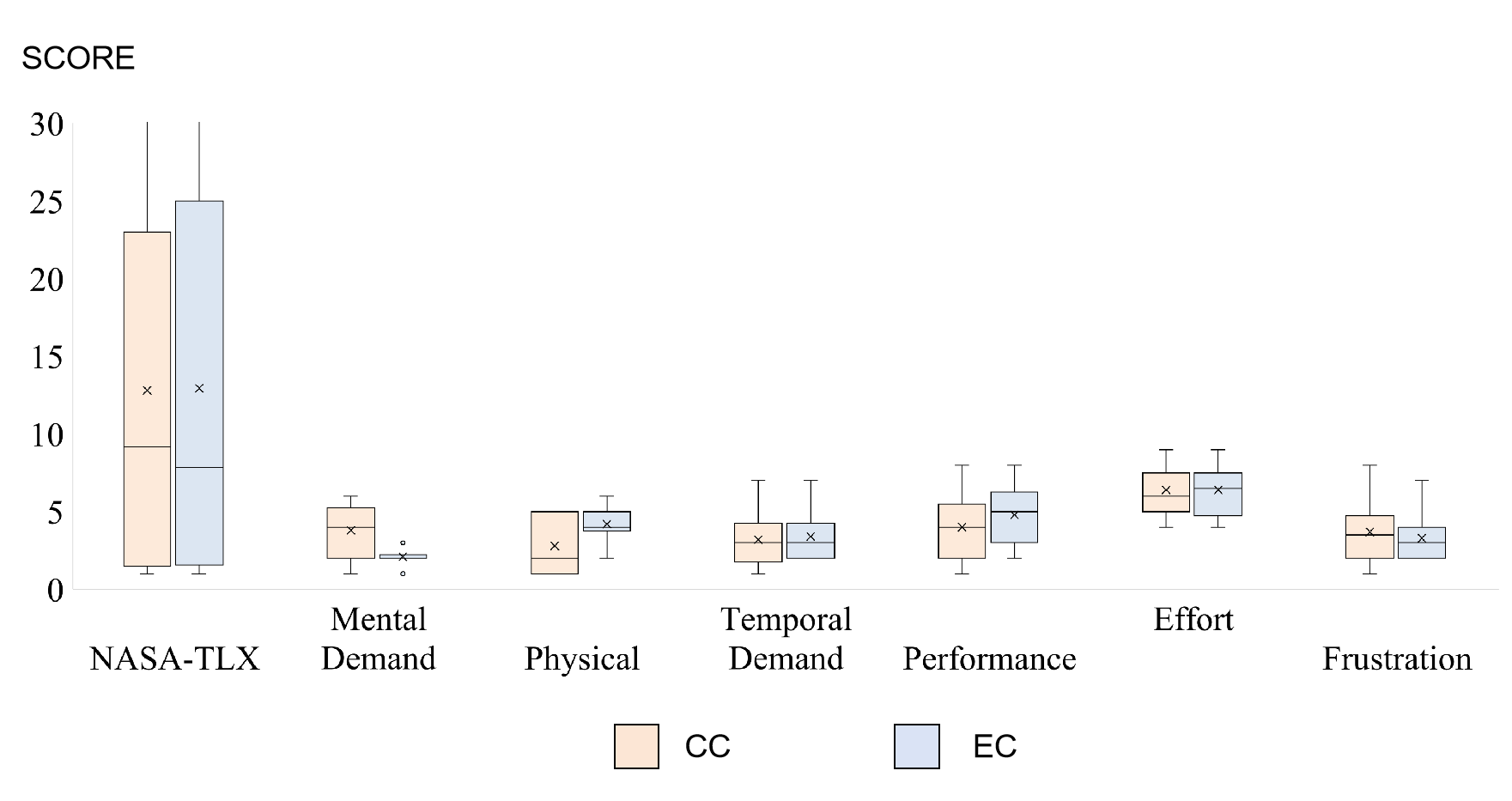}
        \caption{Mean NASA-TLX for Total, Mental Demand, Physical Demand, Temporal Demand, Performance, Effort, or Frustration scores across each method. The colors respectively represent the controller-based locomotion and Embodied CAVE Locomotion.}
        \label{fig:nasa}
\end{figure}

In \autoref{fig:nasa}, the evaluation results of the NASA-TLX scale indicate that compared to controller-based locomotion, action-recognition-based locomotion slightly increased users' cognitive load across multiple dimensions. Specifically, on the dimensions of Total Load and Effort, the scores for action-recognition-based control were noticeably higher than those for controller-based control. This suggests that although the natural interaction method based on body movements is more intuitive and efficient, the walking process not only makes the participants more exhausted than the controller group but also requires more effort.

Although the differences between the $CC$ and $EC$ methods were not significant in terms of physical demand and temporal demand, the scores remained at a low level. The action-recognition-based control received similar or even slightly higher scores on the performance dimension compared to $CC$. This further validates that it can effectively maintain the performance of user tasks while reducing workload. Furthermore, in the Frustration dimension, the score for the action-recognition-based control was also lower than that for $CC$, which means that users experienced less frustration during use.

\subsection{Discussion}

The results of our study provide information on user interaction with the proposed method, particularly in proving the hypothesis.

\textbf{H1}: Compared to  $CC$ , $EC$ can alleviate motion sickness.

The results in \autoref{fig:ssq} support \textbf{H1}. The reasons for the efficiency of our method may be: The fundamental cause of motion sickness lies in the sensory conflict between vestibular signals (from the inner ear) and visual perception, where incongruent motion signals lead to neural mismatch. When users navigate a virtual environment in a CAVE using a traditional controller, their inner ear and proprioceptive system (body position) do not register physical movement, while their visual system perceives motion. This creates a classic sensory conflict. By contrast, the full-body action recognition method employed in this study directly translates the user's physical movements into corresponding virtual locomotion. This creates a more harmonious sensory experience, as the visual motion input aligns consistently with the vestibular and proprioceptive signals generated by the user's own physical effort. Consequently, the observed reduction in SSQ scores is attributable to this enhanced sensory congruence, which minimizes the neural mismatch that triggers motion sickness. The system effectively "tricks" the brain into believing that the perceived visual motion is indeed caused by the user's own physical movement, leading to a more natural and comfortable experience.

\textbf{H2}: The task load of $CC$ is higher than that in $EC$. 

The result in \autoref{fig:nasa} shows that the action-recognition-based embodied interaction can't significantly reduce users' cognitive load. This result disproves the hypothesis that controller-based locomotion imposes a higher task load compared to action-recognition-based methods. The reason might be that, compared to standing still and controlling the scene changes with a controller, stepping in place in the CAVE environment is more physically demanding and requires more effort to control one's pace and direction. However, action-recognition-based control as an embodied interaction method can slightly alleviate users' mental demand and provide a better user experience, offering a superior solution for immersive locomotion in CAVE environments.

\textbf{H3}: The sense of immersion of $EC$ is better than that in $CC$. 

The results in \autoref{fig:ipq} support \textbf{H3}. 
Because participants were constantly moving in the CAVE environment throughout the experiment, controlling their locomotion through their own movements, our method enhanced their sense of presence and immersion in the virtual environment. Additionally, the physical effort required for movement led to a reduced focus on the real-world surroundings outside the virtual scene.

\section{Conclusion, limitation, and future work}

This paper introduces a pipeline for human motion recognition based on locomotion in CAVE that keeps users immersed, significantly reduces motion sickness, and offers a method for immersive locomotion in CAVE environments. By leveraging real-time action classification through our method, this approach achieves natural user interaction in virtual environments, eliminating dependency on handheld controllers. Compared to the $CC$, our method can alleviate motion sickness, improve the sense of immersion. 

However, our method has four limitations. One limitation is that participants must remain within the coverage area of the CAVE cameras. The further they move from the center of the CAVE floor, the less accurate our human pose recognition becomes, particularly in the four corners. In future work, we will expand the tracking area by integrating additional depth sensors or optimizing camera placement to improve robustness in peripheral regions.Another limitation is that users wearing highly complex attire (e.g., fire-resistant suits) may experience degraded performance due to occlusion and limited pose visibility. In future work, we will investigate adaptive algorithms that leverage sparse keypoint estimation or inertial measurement units (IMUs) to compensate for clothing-induced occlusions.
Additionally, our current method requires manual selection of a single user as the primary navigator in multi-user CAVE scenarios, lacking support for collaborative or concurrent recognition. In future work, we will develop multi-agent motion recognition techniques, combining skeletal tracking with identity assignment to enable seamless group interactions.Finally, our approach relies on in-place walking to drive virtual navigation but does not explore how different locomotion modes (e.g., real walking in CAVE vs. in-place stepping) or scaling ratios (e.g., 1:1 vs. redirected walking) or different task durations (e.g., two minutes walking in CAVE vs. ten minutes walking in CAVE) affect user experience. 

In future work, we will introduce redirected walking technology and conduct comparative works to optimize locomotion techniques for immersion, comfort, and spatial awareness.




\bibliographystyle{elsarticle-num} 
\bibliography{main} 
\end{document}